\documentclass[aps,pra,reprint,groupedaddress,superscriptaddress]{revtex4-1}
\usepackage[dvipsnames]{xcolor}
\usepackage{graphicx,amsmath,amssymb}
\usepackage{newtxtext,newtxmath} 
\usepackage{graphicx}
\usepackage[colorlinks,linkcolor=Blue,citecolor=Blue,urlcolor=Blue,breaklinks=true]{hyperref}
\begin{document}
\title{Analog curved spacetimes in the reversed dissipation regime of cavity optomechanics}
\author{F. Bemani}  \email{foroudbemani@gmail.com}
\address{Department of Physics, Faculty of Science, University of Isfahan, Hezar Jerib, 81746-73441, Isfahan, Iran}
\author{R. Roknizadeh} \email{r.roknizadeh@gmail.com}
\author{M. H. Naderi} \email{mhnaderi@phys.ui.ac.ir}
\address{Department of Physics, Faculty of Science, University of Isfahan, Hezar Jerib, 81746-73441, Isfahan, Iran}
\address{Quantum Optics Group, Department of Physics, Faculty of Science, University of Isfahan, Hezar Jerib, 81746-73441, Isfahan, Iran}
\date{\today}

\begin{abstract}
In this paper, we theoretically propose an optomechanical scheme to explore the possibility of simulating the propagation of the collective excitations of the photon fluid in a curved spacetime. For this purpose, we introduce two theoretical models for two-dimensional photon gas in a planar optomechanical microcavity and a two-dimensional array of coupled optomechanical systems. In the reversed dissipation regime (RDR) of cavity optomechanics where the mechanical oscillator reaches equilibrium with its thermal reservoir much faster than the cavity modes, the mechanical degrees of freedom can adiabatically be eliminated. The adiabatic elimination of the mechanical mode provides an effective nonlinear Kerr-type photon-photon interaction. Using the nonlinear Schr\"{o}dinger equation (NLSE), we show that the phase fluctuations in the two-dimensional photon fluid obey the Klein-Gordon equation for a massless scalar field propagating in a curved spacetime. The results reveal that the photon fluid as well as the corresponding metric can be controlled by manipulating the system parameters. 
\end{abstract}

\maketitle
\section{Introduction}
Investigation of an uncontrollable or inaccessible quantum system or quantum systems with a large number of degrees of freedom via some controllable quantum system is known as quantum simulation \cite{Feynman, Lloyd,Georgescu}. To this end, a wide variety of systems such as atoms in optical lattices, trapped ions, nuclear spins, superconducting circuits and photonic systems have been proposed as quantum simulators in different branches of physics such as condensed matter physics, high-energy physics, and cosmology \cite{Georgescu}. 

The general theory of relativity (GTR) is, without doubt, one of the fascinating theories of the history of science describing the most important gravitational phenomena. The light deflection in gravitational  field, gravitational waves, and also appearance of  black holes, are some famous and spectacular predictions of GTR.  One of the most important goals of recent investigations in theoretical physics is the presentation of a unified theory for  gravitation and other quantum fields  by their quantization  on a curved background, i.e., a quantum field theory (QFT) in curved spacetime. This theory has some experimental consequences, such as particle creation, Hawking radiation, etc. (see e.g., \cite{Birrell} and references therein). Even in the case of small curvature, various aspects of QFT in curved spacetime cannot be examined directly with current technologies. Nevertheless, they can be investigated via quantum simulation. The geometric formulation of quantum mechanics is a very important step toward the presentation of the analog models of GTR in quantum  realm. Hence, the investigation of novel models and a better understanding of previous analog models is still open-ended \cite{Novello,Barcelo2005,Unruh2007,Faccio,Sindoni}. Various analog models of GTR have been proposed, such as the acoustic fluid \cite{Unruh}, liquid helium \cite{Volovik,Jacobson}, Fermi gases \cite{Giovanazzi,Giovanazzi2}, slow light \cite{Leonhardt,Leonhardt2,Reznik,Unruh3}, nonlinear electromagnetic waveguides \cite{Schutzhold}, graphene \cite{Cortijo,Vozmediano}, ion rings \cite{Horstmann} and Bose-Einstein condensate (BEC)\cite{Garay,Garay2,Barcelo,Mayoral,Finazzi,Girelli,PAnderson}. The superfluid states of light as a consequence of nonlinearity-induced photon-photon interactions were also used as quantum simulators of quantum field theories on curved spacetimes \cite{Nguyen,Marino,Fouxon,Solnyshkov,Gerace}.

Many of the notions of atomic and molecular physics can be realized on an entirely different scale, namely, in optomechanical systems. The field of optomechanics is currently undergoing rapid experimental and theoretical progress. Typically, electromagnetic radiation and macroscopic mechanical oscillators can interact via radiation pressure in a Fabry-Perot cavity with a movable end mirror. This interaction is the basis of various optomechanical phenomena which can occur in a wide range of system sizes and parameters (for  review, see \cite{Kippenberg,Meystre,Aspelmeyer,Bowen}). Many of the rudiments of quantum optomechanics date back to the early attempts of the investigation of one of the most prominent predictions of GTR, i.e., the gravitational waves. However, during the past decade, significant efforts have been devoted to developing and implementing optomechanical interaction for position or force sensing \cite{Murch,Thompson,Purdy,Motazedifard}, backaction cooling \cite{Teufel,Chan}, quantum state transfer \cite{Palomaki}, optomechanical entanglement generation \cite{Palomaki2,Paternostro,Vitali}, optomechanically induced transparency realization \cite{Zhang,Weis,Hong}, and generating self-sustained mechanical oscillations \cite{Marquardt1}. Another interesting aspect of the optomechanical systems, which has not been investigated much yet, is that they can be considered as promising candidate systems for quantum simulation applications. For instance, it has very recently been shown \cite{Bemani} how to realize the so-called sphere-coherent motional states of a mechanical oscillator in an optomechanical cavity in the presence of a two-level atom and how to use the optomechanical parameters to control the curvature of the sphere.  

In the present paper, we investigate the possibility of simulating the curved spacetimes by photonic fields  interacting with a mechanical oscillator operating in the reversed dissipation regime (RDR) of optomechanics \cite{Nunnenkamp,Toth}. Due to the radiation pressure force, an effective Kerr-type photon-photon interaction arises which is mediated by the mirror vibration. The emergence of fluid-like behavior for photons is the consequence of this photon-photon interaction so that we can use the hydrodynamic equations for the photonic fluid to study the corresponding effective metric for the propagation of fluctuations in the photon fluid. In this manner, the optomechanical system can be regarded as a quantum simulator for the propagation of  the collective excitations of photon fluid in a curved spacetime. We also show that the photonic fluid as well as the corresponding curved spacetime can be controlled  by manipulating the system parameters. 

The rest of the paper is structured as follows.  In Section \ref{sec:Sec2}, we introduce the quantum field description of our first proposed system, i.e., a planar optomchanical microcavity. We discuss the equations of motions in the RDR of optomechanics. In Section \ref{sec:Sec3}, we present our second proposed setup which is a two-dimensional optomechanical array. In Section \ref{sec:Sec4},  we discuss the so-called hydrodynamic version of the nonlinear Schr\"{o}dinger equation (NLSE), which can be linearized to describe an effective metric for the two-dimensional photon fluid. Finally, in Section \ref{sec:Sec5}, we present our concluding remarks as well as some possible outlooks of our work..

\section{ Quantum Field Description of a Planar Optomechanical MicroCavity}\label{sec:Sec2}
\begin{figure}[h!]
	\includegraphics[width=8.5cm]{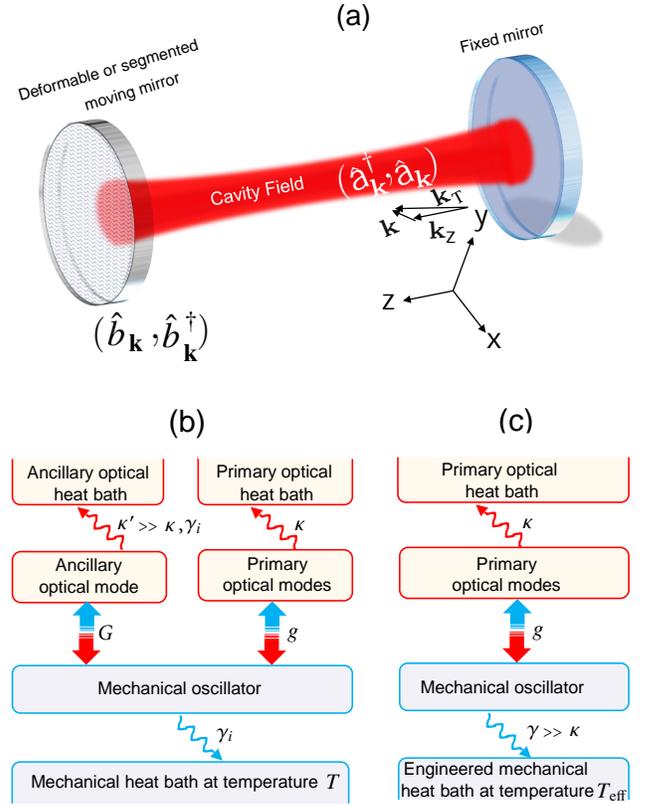}
	\caption{(Color online). Schematic illustration of a Fabry-Perot cavity with a deformable or segmented moving end mirror as the mechanical element interacting with a continuum of optical modes. The mirror spacing is close to, or
		below the dimension of the wavelength of the optical field while the mirror radius of curvature is in the meter regime. (b) The ancillary optical field  with a high damping rate ($ \kappa'\gg \gamma_i, \kappa $) is used to reach the RDR of cavity optomechanics. (c) RDR regime of optomechanics where $ \gamma \gg \kappa $ (see appendix \ref{app1} for details of calculations).}
	\label{Fig:Fig1}
\end{figure}
As the first theoretical proposal, we consider  a planar optomechanical microcavity. As depicted in Fig (\ref{Fig:Fig1}-a), we consider a Fabry-P\'{e}rot cavity with a mechanically deformable or a segmented mirror which can be modeled as a collection of mechanical modes \cite{Klaers3}. In order to describe the system, we separate the total wave vector ${{\bf{k}}_T} =(k_x,k_y,k_z)$, into the transversal in-plane, ${\bf k}= {k_x}{\bf{\hat x}} + {k_y}{\bf{\hat y}}$, and the longitudinal, $k_z$, components. The single photon energy in the limit of paraxial approximation as a function of in-plane and longitudinal wave vectors reads,
\begin{equation}
\hbar {\omega _c}( {\bf k},k_z) = \hbar c\sqrt {{\bf{k}}^2 + k_z^2} ,
\end{equation}
where $c$ denotes the speed of light. In the system under consideration, we assume that the mirror spacing is close to, or below the dimension of the wavelength of the cavity field while the mirror radius of curvature is in the meter regime. Therefore, photons in the $xy$-plane are not confined and one can apply the continuum approximation for the wave vector in the transverse direction. Imposing the boundary conditions on the electromagnetic field inside the optomechanical cavity with curved mirrors of radius $R$ leads to a discrete wave vector in the $z$ direction, namely, ${k_z}\left( r \right) = n\pi /l\left( r \right)$, where the mirrors separation at the distance $r$ from the cavity axis is given by $l\left( r \right) = {l_0} - 2\left( {R - \sqrt {{R^2} - {r^2}} } \right)$. In the limit of paraxial approximation $(r \ll R,|{\bf k}| \ll k_z)$ the photon dispersion relation becomes
\begin{eqnarray}
&&\hbar {\omega _c}({\bf{k}},r) = \hbar c\sqrt {{{\bf{k}}^2} + {{\left( {\frac{{n\pi }}{{l\left( r \right) + z(\bf{r})}}} \right)}^2}} \nonumber\\
&&\qquad\qquad\,\,\,\simeq m{c^2} + \frac{{{\hbar ^2}{\bf{k}}^2}}{{2m}} + {V_{\rm{t}}}(r) - \hbar {g_0}z(\bf{r}), \qquad \qquad \qquad \label{eq:dispersion_relation}
\end{eqnarray}
where $V_{\textrm{t}}(r)=m{\Omega ^2}{r^2}/2$ is the trapping potential, ${g_0} = n\pi c/l_0^2$ is the single-photon optomechanical coupling rate and  $z(\bf{r})$ is the displacement (deformation) of the mirror from its equilibrium position at point $r$.  Therefore, in a Fabry-P\'{e}rot cavity, spatial confinement of the photons in the $z$ direction breaks the symmetry and eventually it provides a finite effective mass (${m} = \hbar n\pi /c{l_0}$) for the photons in the plane perpendicular to the cavity axis \cite{Klaers,Klaers2,Klaers3}. This makes the photon gas in the $xy$-plane to be effectively two-dimensional. Moreover, the mirror curvature provides a trapping potential with frequency $\Omega  = c\sqrt {2/{l_0}R}$. It should be noted that in the case $z({\bf r})=0$ (cavity with fixed mirrors)  we recover the results of Refs. \cite{Klaers,Klaers2,Klaers3}. Moreover, for a rigid non-deformable mirror it does not require to take the mirror displacement to be a function of $\bf{r}$. In such a situation, all optical modes interact with a single mechanical mode.

Here, we consider the case in which the deformable mirror is interacting with a continuum of optical modes, schematically shown in Fig.~(\ref{Fig:Fig1}-a). Therefore, the mirror is considered to be a set of quantum mechanical harmonic oscillators with the same effective masses $m_m$, frequencies $\omega_m$, and the position operator $\hat{z}({\bf r})=z_0 \sqrt{A} [ \hat{b}({\bf r})+\hat{b}^\dag({\bf r})]$ where $\hat{b}({\bf r})$ and $\hat{b}^\dag({\bf r})$ satisfy the commutation relation $[\hat{b}({\bf r}),\hat{b}^\dag({\bf r}') ]= \delta({\bf r}-{\bf r}')$, and $[ {\hat b}({\bf r}),{\hat b}({\bf r}') ] = 0 = [ \hat b^\dag({\bf r}) ,\hat b^\dag ({\bf r}') ]$, $z_0=\sqrt{\hbar/2m_m \omega_m}$ is the zero-point fluctuation of the mirror's position and $A$ is the effective area of the mechanical modes. By definition the operator $\hat n_b(t,{\bf{r}}) \equiv {{\hat b }^\dag }(t,{\bf{r}})\hat b (t,{\bf{r}}) $ can be interpreted as a sort of two-dimensional phonon density. The Fourier transform of $\hat{a}_{\bf{k}}$ ($\hat{a}^\dag _{\bf{k}}$) is defined to be the two-dimensional cavity photon field operator $\hat{\Psi}(t,\bf r)$  ($\hat{\Psi}^\dag (t,\bf r)$)
\begin{eqnarray}
\hat \Psi (t,{\bf{r}}) = \int\limits {\frac{{d^2{\bf{ k}}}}{{{{(2\pi )}^2}}}} {{\hat a}_{{\bf{ k}}}}{e^{i{\bf{ k}}.{\bf{r}}}}\, .
\end{eqnarray}
The field operators satisfy the following equal-time commutation relations $ [ {{{\hat \Psi } }(t,{\bf{r}}),\hat \Psi ^\dag (t,{\bf{r'}})} ] = \delta ({\bf{r}} - {\bf{r'}}) $ and $ [ {\hat \Psi (t,{\bf{r}}),\hat \Psi (t,{\bf{r'}})}] = 0 = [ {{{\hat \Psi }^\dag }(t,{\bf{r}}),{{\hat \Psi }^\dag }(t,{\bf{r'}})}] $.
The continuum of optical modes, characterized by the transverse wave vector $\textbf{k}$ and the annihilation (creation) operator $a_{\bf k}$ ($a_{\bf k}^\dag$), is coupled to the vibrating mirror via the radiation pressure coupling. The optical field operators satisfy the commutation relations $[ {\hat a_{\bf{k}} ,{{\hat a}_{{\bf{k'}}}}^\dag} ] = {(2\pi )^2}\delta ({\bf{k}} - {\bf{k'}})$ and $\left[ {{{\hat a}_{\bf{k}}},{{\hat a}_{{\bf{k'}}}}} \right] = 0 = [ {\hat a_{\bf{k}}^\dag ,\hat a_{{\bf{k'}}}^\dag } ]$. In other words, the cavity free spectral range is much larger than the mechanical frequency (the single-longitudinal-mode assumption).  We restrict our considerations for the primary mode to the case of a single-longitudinal-cavity mode. Using the dispersion relation of Eq.~(\ref{eq:dispersion_relation}), the many-body Hamiltonian of the system is expressed as 
\begin{eqnarray}
&&\hat H/ \hbar  = \int {{d^2}{\bf{r}}}  {{\hat \Psi }^\dag }(t,{\bf{r}})\left[ {{\Delta}  - g\left( {\hat b(t,{\bf{r}}) + {{\hat b}^\dag }(t,{\bf{r}})} \right)} \right]{{\hat \Psi } }(t,{\bf{r}}) \nonumber \\
&& \qquad + \int {{d^2}{\bf{r}}}{\omega _m}{{\hat b}^\dag }(t,{\bf{r}})\hat b(t,{\bf{r}}) ,
\label{Eq:Hamiltonian1}
\end{eqnarray}
where we have defined $\Delta = mc^2/\hbar -\hbar \nabla ^2/2m+{V_{\rm{t}}}(r)$. 
This Hamiltonian consists of the free energy of the mechanical modes and the optical modes, the trapping potential due to the mirror curvature, and the optomechanical interaction energy. We have also defined $g=g_0z_0 \sqrt{A}$ as the effective optomechanical coupling strength. 

We investigate the dynamics of the system by the Heisenberg-Langevin equations associated with the Hamiltonian of Eq. (\ref{Eq:Hamiltonian1})
\begin{subequations}
	\begin{equation}
	{\partial _t}\hat \Psi (t,{\bf{r}}) =  - i\left[ {{\Delta }- g\left( {\hat b(t,{\bf{r}}) + {{\hat b}^\dag }(t,{\bf{r}})} \right)} \right]\hat \Psi (t,{\bf{r}}),
	\label{Eq:HeisenbergLangevin1}
	\end{equation}
	\begin{equation}
	{\partial _t}\hat b(t,{\bf{r}}) \!=\!  - i\left( {{\omega _m} \!- i\gamma /2} \right)\hat b(t,{\bf{r}}) \!+ ig{{\hat \Psi }^\dag }(t,{\bf{r}})\hat \Psi (t,{\bf{r}}) \!+\! \sqrt \gamma  {{\hat b}^{in}}(t,{\bf{r}}),
	\label{Eq:HeisenbergLangevin2}
	\end{equation}
\end{subequations}
where the zero-mean operator ${{\hat b}^{in}}(t,{\bf{r}})$ that denotes the mechanical noise operator satisfies the commutation relation $ [{{\hat b}^{in}}(t,{\bf{r}}),{{\hat b}^{in,\dag }}(t',{\bf{r}})] = \delta (t - t')$ and the second order correlation $ \langle {{\hat b}^{in,\dag }}(t,{\bf{r}}){{\hat b}^{in}}(t',{\bf{r}})\rangle  = {{\bar n}_{\rm{f}}}\delta (t - t'), $ in which we have assumed that the cavity is in zero temperature and $\bar{n}_{\rm{f}} = {\left( {\exp \left( {\hbar {\omega_m}/{k_B}T_{\rm{eff}}} \right) - 1} \right)^{ - 1}}$ is the mean number of thermal phonons of the mechanical oscillators at heat bath temperature $T_{\rm{eff}}$, with $k_B$ being the Boltzmann constant. Here, we assume that the fluctuation-dissipation process affects only the mechanical modes. This assumption is justified by using a cavity with a high-quality factor (small damping rate $\kappa$) and a mechanical oscillator with a large damping rate. Of course, it is not the normal dissipation regime of cavity optomechanics where the mechanical dissipation rate $\gamma$ is much smaller than the cavity linewidth $ \kappa$.  Therefore, one has to engineer mechanical modes with an effective large damping rate. The mechanism of realizing RDR of optomechanics is depicted in Figs. (\ref{Fig:Fig1}-b) and (\ref{Fig:Fig1}-c). The detailed considerations for the realization of the RDR in the system under study are descibed in Appendix \ref{app1}. Briefly, RDR with respect to the optical mode $ \hat{a}_{\bf{k}}$ can be achieved by optomechanical sideband cooling of a high-Q mechanical mode with an ancillary optical mode possessing a large damping rate $ \kappa'\gg\kappa $. In principle, for time scales longer than $\gamma^{-1}$ and much shorter than $\kappa^{-1}$ we conclude that the photonic subsystem becomes isolated from the environment. Based on the calculations in Appendix \ref{app1}, for a typical optomchanical system with a mirror with intrinsic frequency $\omega_i = 2\pi \times 10 \,\,\rm{MHz}$ interacting with an ancillary optical mode with damping rate $\kappa' = 12.5 \,\,\rm{MHz} $, it is possible to increase the mechanical damping rate to $\gamma$ up to $10\,\,\rm{MHz}$.

Since the mechanical damping rate is much greater than the cavity decay rate ($\gamma \gg \kappa$), the mechanical mode can be adiabatically eliminated on time scales greater than $\gamma^{-1}$. For this purpose, one can formally integrate Eq.~(\ref{Eq:HeisenbergLangevin2}) to obtain
\begin{eqnarray}
&&\hat b(t,{\bf{r}}) = {{\hat b}}(0,{\bf{r}}){e^{ - i\left( {{\omega _m} - i\gamma /2} \right)t}} + ig\int\limits_0^t {dt'\hat n(t',{\bf{r}}){e^{i\left( {{\omega _m} - i\gamma /2} \right)(t' - t)}}}  \nonumber\\
&&\qquad\qquad\qquad\qquad+ \sqrt \gamma  \int\limits_0^t {dt'{e^{i\left( {{\omega _m} - i\gamma /2} \right)\left( {t' - t} \right)}}{{\hat b}^{in}}(t',{\bf{r}})} \,\qquad 
\end{eqnarray}
where we have defined $ \hat n(t,{\bf{r}}) \equiv {{\hat \Psi }^\dag }(t,{\bf{r}})\hat \Psi (t,{\bf{r}}) $ as the two-dimensional photon density. For time scales long compared to $\gamma^{-1}$, the first term becomes zero. Substituting this equation into Eq. (\ref{Eq:HeisenbergLangevin1}) the following equation is obtained for the optical mode
\begin{eqnarray}
{\partial _t}\hat \Psi (t,{\bf{r}}) =  - i{\Delta }  \hat \Psi (t,{\bf{r}}) + {{\hat f}^{in}}(t,{\bf{r}}) \qquad\qquad\qquad\qquad\qquad \nonumber \\
\qquad \quad  - 2i{g^2}\left[ {\int\limits_0^t {dt'{e^{\gamma (t' - t)/2}}\sin {\omega _m}(t' - t)\hat n(t',{\bf{r}})} } \right]\hat \Psi (t,{\bf{r}}) , 
\label{Eq:HeisenbergLangevin3}
\end{eqnarray}
where we have introduced the generalized noise operator 
\begin{equation}
{{\hat f}^{in}}(t,{\bf{r}}) = \sqrt \gamma  g\hat \Psi (t,{\bf{r}})\int\limits_0^t  dt'\left[ {{e^{i\left( {{\omega _m} - i\gamma /2} \right)\left( {t' - t} \right)}}{{\hat b}^{in}}(t',{\bf{r}}) + H.c.} \right] .
\end{equation}
For time scales much longer than the mechanical characteristic time $\gamma^{-1}$ this generalized noise operator can be approximated by ${\hat f}^{in}(t,{\bf{r}}) \simeq 0$.  The net effect of the mirror is a redistribution of the photons between the various transversal modes. Now, in the RDR the mirror motion can be adiabatically eliminated, resulting in a mechanical field that is affected too much by the optical field operators. Using this approximation and defining 
\begin{equation}
\mathcal{T}\left( t \right) = -\int\limits_0^t {dt'{e^{  -\gamma \left( {t - t'} \right)/2}}\sin { \omega _m}\left( {t' - t} \right)}\, ,
\end{equation}
which for time scales much longer than $\gamma_{\rm{}}^{-1}$ can be approximated as
\begin{equation}
\mathcal{T} \equiv \mathcal{T} (t \gg {\gamma ^{ - 1}}) \simeq \frac{\omega _m}{\gamma ^2/4 +  \omega _m^2}\, ,
\end{equation}
Eq. (\ref{Eq:HeisenbergLangevin3}) becomes
\begin{equation}
{\partial _t}\hat \Psi (t,{\bf{r}}) =  - i\left[ {\Delta- 2{g^2}{\cal T}\hat n(t,{\bf{r}})} \right]\hat \Psi (t,{\bf{r}}).
\label{Eq:HeisenbergLangevin4}
\end{equation}
Altogether, the adiabatic elimination of the mirror motion has two consequences on the dynamics of the system. First, it provides a reservoir for the photons inside the cavity that can exchange excitations. Second, it gives rise to a nonlinear Kerr-type photon-photon interaction with a modified interaction rate, $2g^2\mathcal{T}$. 

We now apply the mean field approximation in which the quantum operator is separated into $\hat \Psi (t,{\bf{r}}) = {\Psi _0}(t,{\bf{r}}) + {{\hat \chi }}\left( {t,{\bf{r}}} \right)$
where $ {\Psi _0}(t,{\bf{r}})$ is the mean field, and $\hat \chi ( t,{\bf{r}})$ is the fluctuation operator with a
zero mean value around the mean field. For subsequent mathematical convenience, we choose $ {{\hat \chi }}\left( {t,{\bf{r}}} \right) \equiv {\Psi _0}(t,{\bf{r}})\hat \phi \left( {t,{\bf{r}}} \right) $. The classical mean field and the fluctuation operator satisfy, respectively	
\begin{subequations}
	\begin{eqnarray}
	&&i\hbar {\partial _t}{\Psi _0} = \left[ { - \frac{{{\hbar ^2}}}{{2m}}{\nabla ^2} + \tilde{V}_{\rm{t}}(r) + \mathcal{G}|{\Psi _0}{|^2}} \right]{\mkern 1mu} {\Psi _0},
	\label{Eq:HeisenbergLangevin5}
	\\
	&&i\hbar {\partial _t}\hat \phi  =  - \left[ {\frac{{{\hbar ^2}}}{{2m}}{\nabla ^2} + \frac{{{\hbar ^2}}}{m}\frac{{\nabla {\Psi _0}}}{{{\Psi _0}}}\nabla } \right]\hat \phi  + n\mathcal{G}(\hat \phi  + {{\hat \phi }^\dag }), \qquad
	\label{Eq:HeisenbergLangevin6}
	\end{eqnarray}
\end{subequations}
Equation (\ref{Eq:HeisenbergLangevin6}) is the analog of the Bogoliubov-de Gennes equation for the fluctuations.  We also have defined $\tilde{V}_{\rm{t}}(r)=mc^2+{V_{{\rm{t}}}}(r)$ , and the photon-photon coupling strength $\mathcal{G}$   
\begin{equation}
\mathcal{G}=-2\hbar g^2 \mathcal{T}= -2\hbar \omega _m \frac{ g^2 }{\gamma ^2 +  \omega _m^2}\,,
\end{equation}
is determined by the optomechanical parameters.
\section{Array of coupled optomechanical systems}\label{sec:Sec3}
\begin{figure}
	\includegraphics[width=7cm]{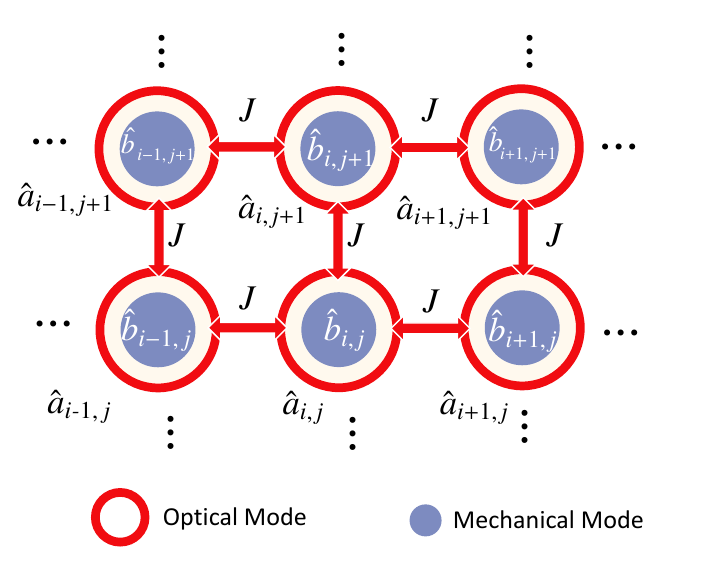}
	\caption{(Color online). Schematic illustration of a two-dimensional array of coupled optomechanical systems.}
	\label{Fig:Fig5}
\end{figure}
As the second theoretical proposal, we consider a two-dimensional array of coupled optomechanical systems schematically depicted in Fig~(\ref{Fig:Fig5}). This system could be realized experimentally in optomechanical crystals. One- and two-dimensional optomechanical crystals have been recently realized experimentally \cite{Safavi,Chan2016}. In this case, optical (mechanical) modes on each site interact locally via radiation pressure with rate $g'$while photons can tunnel between adjacent sites with hopping rate $J$. Since in such experiments phonon tunneling rate is much smaller (four orders of magnitude) than the photon tunneling rate, we can neglect phonon hopping in our considerations.  The optomechanical array Hamiltonian is given by
\cite{Ludwig,Chen,Peano} 
\begin{eqnarray}
&&	\hat H/\hbar  = \sum\limits_{i,j} {\left[ {{\omega _c}\hat a_{ij}^\dag {{\hat a}_{ij}} + {\omega _m}\hat b_{ij}^\dag {{\hat b}_{ij}} - {g'}\hat a_{ij}^\dag {{\hat a}_{ij}}(\hat b_{ij}^\dag  + {{\hat b}_{ij}})} \right]} \nonumber \\ && \qquad\qquad\qquad\qquad \qquad\quad  +\sum\limits_{\left\langle {i,j,k} \right\rangle } {J\left( {\hat a_{ij}^\dag {{\hat a}_{kj}} + \hat a_{ij}^\dag {{\hat a}_{ik}}} \right)} \, ,
\label{Eq:Optomechanical_Array_Hamiltonian}
\end{eqnarray}
where ${\hat {b}}_{ij}$ and ${\hat {a}}_{ij}$, respectively, stands for the bosonic annihilation operators for phonon and photons on the lattice site denoted by $ (i,j) $. In  the last term $\left\langle {i,j,k} \right\rangle$ denotes the summation over all adjacent lattice sites. Similar to the case of planar optomechanical microcavity, we consider the system in the RDR and write down the Heisenberg-Langevin equations for the quantum fields
\begin{subequations}
	\begin{eqnarray}
	&&{\partial _t}{{\hat a}_{ij}} =  - \left( {i{\omega _c} + \kappa } \right){{\hat a}_{ij}} + i{g'}(\hat b_{ij}^\dag  + {{\hat b}_{ij}}){{\hat a}_{ij}}\nonumber\\
	&&\qquad\quad\quad\quad- iJ\left( {{{\hat a}_{i - 1,j}} + {{\hat a}_{i + 1,j}} + {{\hat a}_{i,j + 1}} + {{\hat a}_{i,j - 1}}} \right)\,,  \label{Eq:Heisenberg_Equation1}
	\end{eqnarray}
	\begin{eqnarray}
	&&	{\partial _t}{{\hat b}_{ij}} =  - \left( {i{\omega _m} + \gamma } \right){{\hat b}_{ij}} + i{g'}\hat a_{ij}^\dag {{\hat a}_{ij}} - \sqrt \gamma  \hat b_{ij}^{in}(t)\,.
	\label{Eq:Heisenberg_Equation2}
	\end{eqnarray}
\end{subequations}
 We now take the field operators to be continuous functions of $ x=ih $ and $ y=jh $ where $ h $ is the cell spacing, i.e., $
{{\hat a}_{ij}}\left( t \right) \to  \hat \Psi (t,{\bf{r}})$ and ${{\hat b}_{ij}}\left( t \right) \to  \hat b(t,{\bf{r}})$. Here, a remark is in order to elucidate the validity of the continuum approximation. For the sake of simplicity, one may consider Eq.~(\ref{Eq:Heisenberg_Equation1}) when $ g'=0 $ and the dissipation is neglected. Assuming $\alpha_{i,j}=\exp[{{\rm i}( k_i i+k_j j)-{\rm i}\omega t}] $ where $ \alpha_{i,j}=\langle \hat{a}_{ij}\rangle$ one can determine the dispersion relation. Moreover, $ k_i $ and $ k_j $ are the wave vector of the collective mode and $\omega$ is the corresponding frequency. Inserting this assumption into Eq.~(\ref{Eq:Heisenberg_Equation1}), we get the dispersion relation $ \omega  = {\omega _c} + 2J{\left[ {cos({{k_i}}) + cos({{k_j}}) } \right]} $. Now, for  a slowly varying optical field which is justified when the optical wave vector $ k_i,k_j\ll1 $, we can use the continuum approximation in Eqs.~(\ref{Eq:Heisenberg_Equation1}) and (\ref{Eq:Heisenberg_Equation2}). In the continuum representation the slowly varying approximation becomes $|{\bf{k}}|=\sqrt{k_x^2+k_y^2}\ll 1/h $. We should mention that $k_x$ and $k_y$ are the continuum representations of $ k_i $ and $ k_j $, respectively.
Under the continuum approximation, the equations of motion take the following forms
\begin{subequations}
	\begin{eqnarray}
	{\partial _t}\hat \Psi (t,{\bf{r}}) =  - \left( {i{\omega _c} + \kappa } \right)\hat \Psi (t,{\bf{r}}) + i{g'}\left[ {{{\hat b}^\dag }(t,{\bf{r}}) + \hat b(t,{\bf{r}})} \right]\hat \Psi (t,{\bf{r}})\nonumber\\
 - iJ\left[{\hat \Psi (t,{\bf{r}} \!-\! h\hat x) \!+\! \hat \Psi (t,{\bf{r}} \!+\! h\hat x) \!+\! \hat \Psi (t,{\bf{r}} \!-\! h\hat y) \!+\! \hat \Psi (t,{\bf{r}} \!+\! h\hat y)} \right]\,, \nonumber\\	
	\label{Eq:Array_Equation_of_motion1}
	\end{eqnarray}
	\begin{equation}
	{\partial _t}\hat b(t,{\bf{r}}) =  - \left( {i{\omega _m} + \gamma } \right)\hat b(t,{\bf{r}}) + i{g'}\hat n(t,{\bf{r}}) - \sqrt \gamma  {{\hat b}^{in}}(t,{\bf{r}})\,.
	\label{Eq:Array_Equation_of_motion2}
	\end{equation}
\end{subequations}
We now expand the field operators $\hat \Psi (t,{\bf{r}} \pm h\hat x)$ and $ \hat \Psi (t,{\bf{r}} \pm h\hat y)  $ as Taylor series  up to second order in $h$
\begin{eqnarray}
\hat \Psi (t,{\bf{r}} \pm h\hat x) \cong \hat \Psi (t,{\bf{r}}) \pm h\frac{{\partial \hat \Psi (t,{\bf{r}})}}{{\partial x}} + \frac{{{h^2}}}{2}\frac{{{\partial ^2}\hat \Psi (t,{\bf{r}})}}{{\partial {x^2}}}\, ,\\
\hat \Psi (t,{\bf{r}} \pm h\hat y)\cong \hat \Psi (t,{\bf{r}}) \pm h\frac{{\partial \hat \Psi (t,{\bf{r}})}}{{\partial y}} + \frac{{{h^2}}}{2}\frac{{{\partial ^2}\hat \Psi (t,{\bf{r}})}}{{\partial {y^2}}}\, .
\label{Eq:TaylorSeries}
\end{eqnarray}	
Under this approximation, Eq.~(\ref{Eq:Array_Equation_of_motion1}) takes the form
\begin{eqnarray}
&&{\partial _t}\hat \Psi (t,{\bf{r}}) =  - iJ{h^2}{\nabla ^2}\hat \Psi (t,{\bf{r}}) - \left( {i{\omega _c} + \kappa  + 4iJ} \right)\hat \Psi (t,{\bf{r}}) \nonumber\\
&&\qquad\qquad \qquad\quad\quad\quad+ i{g'}\left[ {{{\hat b}^\dag }(t,{\bf{r}}) + \hat b(t,{\bf{r}})} \right]\hat \Psi (t,{\bf{r}}) \, .
\label{Eq:Array_Equation_of_motion3}
\end{eqnarray}	
Again, applying the adiabatic approximation for the mechanical modes, we arrive at a NLSE describing the two-dimensional photon field in the optomechanical array
\begin{equation}
\label{eq:GPE}
i\hbar {\partial _t}\hat \Psi\left( {t,{\bf{r}}} \right)  = \left[ { - \frac{{{\hbar ^2}}}{{2m}}{\nabla ^2} + \tilde{V}_{\rm{t}}(r) +\mathcal{G}'|\hat \Psi \left( {t,{\bf{r}}} \right) {|^2}} \right]{\mkern 1mu} \hat \Psi \left( {t,{\bf{r}}} \right) ,
\end{equation}
with $ m = \hbar /2J{h^2}$, ${\mathcal{G}}' = -2\hbar {g'^2}\mathcal{T} $ and $\tilde V_{\rm{t}} = \hbar \left( {{\omega _c} + 4J} \right) $. We should notice that in contrast to the former case, here photons gain mass due to the hopping between lattice sites. Moreover, the potential $ \tilde V_{\rm{t}}$ could be generally space dependent by introducing a space dependent optical frequency $ \omega _c(\bf{r}) $ or hopping rate $  J(\bf{r}) $.
Using the mean-field approximation, the classical mean field and the fluctuation operator satisfy,  respectively, similar equations as Eqs.~(\ref{Eq:HeisenbergLangevin5}) and (\ref{Eq:HeisenbergLangevin6}).
\section{The effective metric for the propagation of the fluctuations in the photonic fluid}\label{sec:Sec4}	
\begin{figure}
	\begin{center}
		\includegraphics[width=9cm]{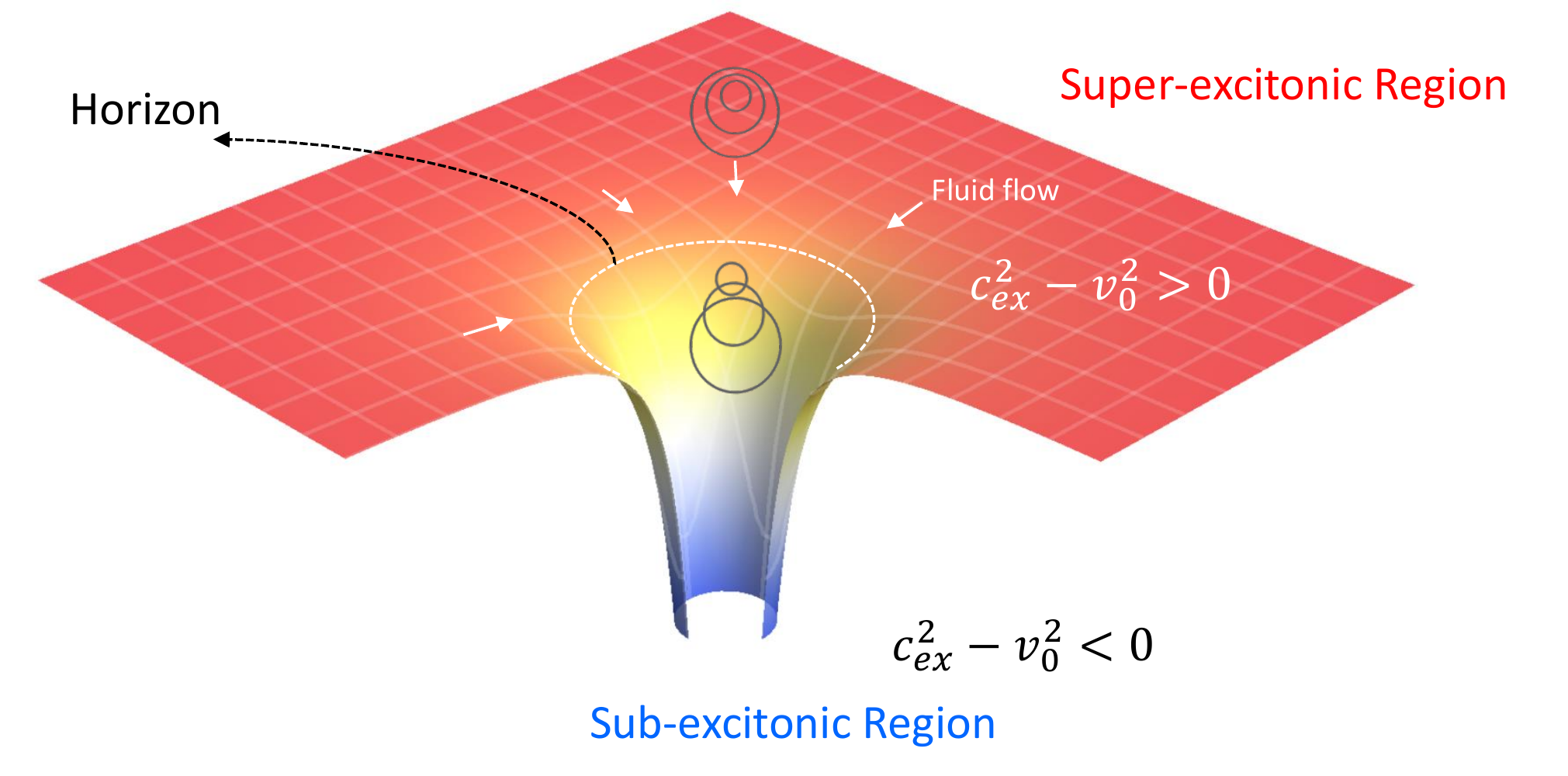} 
	\end{center}
	\caption{(Color online). Schematic diagram of a curved analog spacetime with an analog black hole.} 
	\label{Fig:Fig4}
\end{figure}

In order to  interpret clearly Eqs.~(\ref{Eq:HeisenbergLangevin5}) and (\ref{Eq:HeisenbergLangevin6}), we write the quantum field operator, $\hat\Psi$, in the so-called Madulang representation, namely
\begin{eqnarray}
&& \Psi_0 = \sqrt{n}\- e^{i\theta}\ , \nonumber \\ && \hat\Psi = \sqrt{n+\delta \hat{n}}\  e^{i(\theta + \delta \hat{\theta})}\simeq \Psi_0 (1+ \frac{\delta \hat{n}}{2n} +i\delta \hat{\theta} )\ ,
\end{eqnarray}
where  ${\delta {{\hat n}}}$ and $\delta \hat{\theta}$ denote, respectively, the amplitude of the photon number density fluctuation and the phase fluctuation.
The equations of motion for the density fluctuation and the phase fluctuation read
\begin{subequations}
	\begin{equation}
	{\partial _t}\delta \hat n =  - \nabla ({{\bf{v}}_0}\delta \hat n + \frac{{\hbar n}}{m}\nabla \delta \hat \theta ), \label{Eq:hydrodynamics2}
	\end{equation}
	\begin{equation}
	\hbar {\partial _t}\delta \hat \theta  \!=\!  - \hbar {{\bf{v}}_0}\nabla \delta \hat \theta \! -\! \frac{{mc_{\rm{ex}}^2}}{n}\delta \hat n \!+\! \frac{{mc_{\rm{ex}}^2}}{{4n}}{\xi ^2}\nabla [n\nabla (\frac{{\delta \hat n}}{n})], \label{Eq:hydrodynamics1}
	\end{equation}
\end{subequations}
where ${{\bf{v}}_0} = \hbar \nabla \theta /m$ and $c_{\rm{ex}}=\sqrt{n\mathcal{G}/m}$ are the local velocity of fluid and the local speed of excitations, respectively. The so-called healing length is defined to be $\xi\equiv 1/mc_{\rm{ex}}$. 
Within the hydrodynamic approximation, i.e., over the length scales much larger than $\xi$
the last term in Eq.(\ref{Eq:hydrodynamics1}) can be safely ignored and thus
\begin{equation}
\delta \hat n \simeq  - \frac{\hbar n}{{mc_{\rm{ex}}^2}}\left[ {{{\bf{v}}_0}\nabla \delta \hat \theta  + {\partial _t}\delta \hat \theta } \right].
\label{Eq:hydrodynamics3}
\end{equation}
Combining Eq. (\ref{Eq:hydrodynamics3}) and Eq. (\ref{Eq:hydrodynamics2}) results in 
\begin{equation}\label{eqfase}
- ({\partial _t} + \nabla {{\bf{v}}_0})\frac{n}{{m{c_{\rm{ex}}^2}}}({\partial _t} + {{\bf{v}}_0}\nabla )\delta \hat \theta  + \nabla \frac{n}{m}\nabla \delta \hat \theta  = 0\;.
\end{equation}
These fluctuations, within the hydrodynamic approximation, are analogous to the collective equantum field on a curved metric. In fact, the photon phase fluctuation obeys the covariant Klein-Gordon equation for a massless scalar field propagating in a curved spacetime 
\begin{equation}
\Box \delta \hat \theta  = 0\;,
\end{equation}
where the d'Alembertian operator depends on an effective metric, $g^{\mu\nu}$, and is given by 
\begin{equation}\label{box}
\Box=\frac{1}{\sqrt{-g}}\partial_\mu (\sqrt{-g}g^{\mu\nu}\partial_\nu ) \;,
\end{equation}
with
\begin{equation}
{g_{\mu \nu }} = \frac{n}{{m{c_s}}}\left[ {\begin{array}{*{20}{c}}
	{ - \left( {{c_{\rm{ex}}^2} - {{\bf{v}}_0}.{{\bf{v}}_0}} \right)}&{ - v_0^i}\\
	{ - v_0^j}&{{\delta _{ij}}}
	\end{array}} \right].
\label{Eq:Metric}
\end{equation}
Here, the Greek indices range from $0$ to $2$, $(i,j)$ range from $1$ to $2$, and $g$ is the determinant of $g_{\mu \nu}$.
The line element of this spacetime is given by
\begin{equation}
\label{lelm}
d{s^2}\! =\! {g_{\mu \nu }}d{x^\mu }d{x^\nu } \!=\! \frac{n}{{m c_{\rm{ex}}}}\left[ { \!-\! {c_{\rm{ex}}^2}d{t^2} \!+\! (d{\bf{r}}\! -\! {{\bf{v}}_0}dt)(d{\bf{r}}\! -\! {{\bf{v}}_0}dt)} \right].
\end{equation}
This derivation is similar to the derivation of analog spacetime given in \cite{Garay,Garay2,Barcelo,Mayoral,PAnderson}. A singularity may occur if the flow velocity and the local excitation velocity have equal values. As depicted in Fig.~(\ref{Fig:Fig4}), this configuration corresponds to a black hole since excitation waves traveling with $c_{\rm{ex}}<v_0$ are trapped inside the \textit{superexcitonic} region and they are not able to propagate backward. Although analog models of GTR have a limited ability to simulate all aspects of GR, they provide accessible experimental models for quantum field theory in curved spacetime. Some properties (mainly kinematic) of GTR can be investigated by the analogies with accessible physical systems.  

According to Eqs.~(\ref{Eq:gammaomega}) and (\ref{Eq:FrequencyShift}) if $\omega_{\rm{opt}}>\omega_{i}$, where $ \omega_{\rm{opt}} $ is the mechanical frequency shift due to radiation pressure, then $\mathcal{G}$ will be a positive parameter but it is  physically unacceptable solution because the  system enters an unstable region. Therefore, the parameter $\mathcal{G}$ can only have negative values for a bare planar optomechanical microcavity. This is equivalent to the imaginary velocity for excitation waves and hence they feel a Euclidean metric. One may introduce positive nonlinearities into the system by placing an extra Kerr medium inside the optomechanical microcavity \cite{Shahidani1,Shahidani2} to achieve a real velocity of excitations. For the array of coupled optomechanical systems both the hopping rate $ J $  (effective mass of the photon) \cite{Ludwig,Chen,Peano} and the parameter $\mathcal{G}'$ have negative values and hence velocity of excitations in this case is a real parameter. This is equivalent to the real excitonic speed and, consequently, correponds to a Lorentzian metric for the analog spacetime which is suitable for the investigation of analog black holes. In both cases, the velocity of cavity excitations can be controlled by the ancillary optical mode.   

The real and imaginary values of the speed of excitations correspond, respectively, to the repulsive and attractive
photon-photon interactions. It is important to note that the speed of excitations is not a real speed which can take imaginary values. The physics behind the problem becomes more clear if we describe the system by the differential geometry, i.e., attribute a metric to these different situations. In the case where there is an attractive interaction between photons the local speed of the sound becomes imaginary and hence the fluctuations feel a Euclidean metric. In the classical fluid dynamics however the speed of sound is defined as $ {c^2} = {{\partial p}}/{{\partial \rho }} $, where $ \rho $  is the density of the material and $ p$  is the pressure. We can also define the compressibility as $ \beta  = {1}/{{\rho {c^2}}} $. As we expect, in the usual case in classical hydrodynamics the compressibility has a positive value which means an increase in pressure induces a reduction in volume. Therefore, one should revisit the classical notion of the compressibility in the case of the fluid description of the attractive photons: the system of attractive photons can be understood as a fluid with negative compressibility. 
\section{ Conclusions}\label{sec:Sec5}
\begin{figure}
	\includegraphics[width=8.5cm]{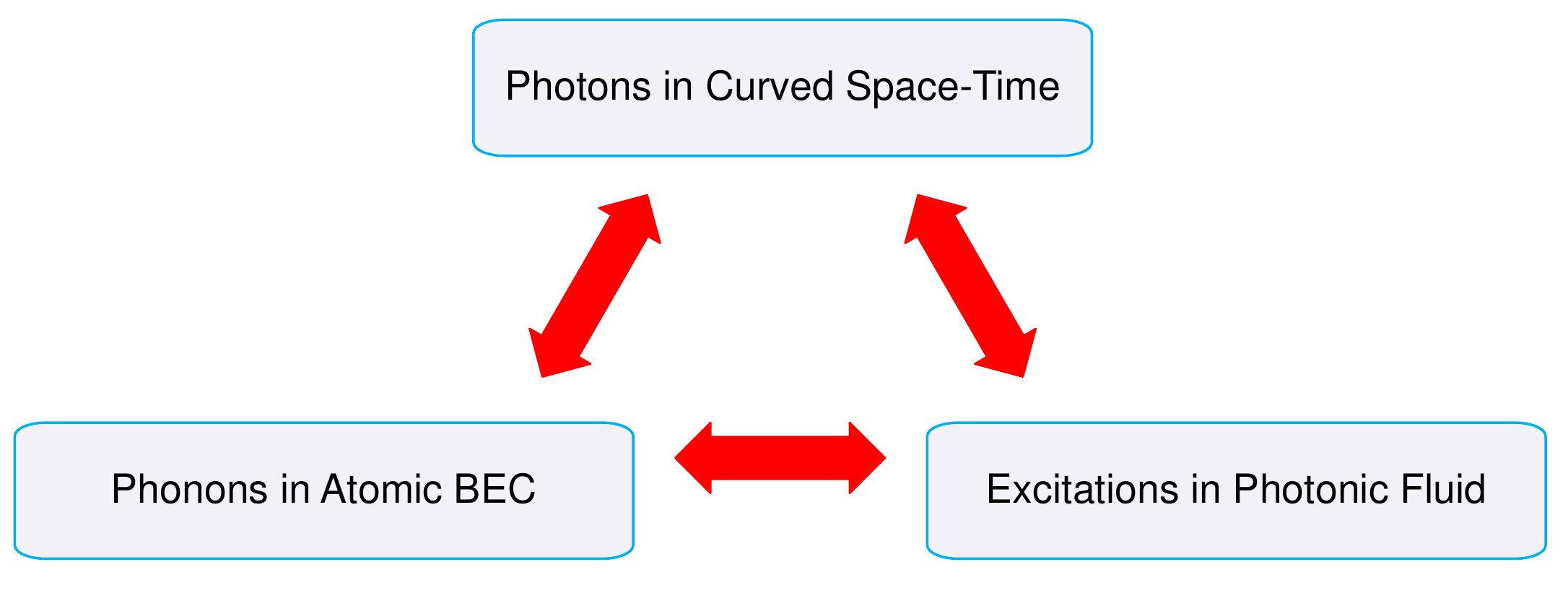}
	\caption{(color online). Schematic diagram of the analogies that have been
		used in this paper.}
	\label{Fig:Fig6}
\end{figure}
In summary, we have introduced a theoretical scheme for the quantum simulation of the curved spacetimes in two types of optomechanical systems operating in the RDR , that is a planar optomechanical microcavity and 
a two-dimensional array of coupled optomechanical systems. Such optomechanical systems are realizable with the state-of-the-art technologies. Schematic diagram of the analogies that have been
used in the paper is presented in Fig~(\ref{Fig:Fig6}). As pointed out in the introduction BECs have been proposed as the simulators of the curved spacetimes. In this paper, we formulate optical fields interacting with optomechanical systems which have a close analogy with BECs as the simulators of the curved spacetime. The main advantage of the current investigation, compared with those carried out previously and cited in the introduction, is the variety in the system sizes, parameters and configurations. RDR of cavity optomechanics is not a situation that is ubiquitous in optomechanics. Although a dissipative quantum reservoir for microwave photons in a superconducting circuit using a mechanical oscillator is realized \cite{Toth}, however, we believe that our work can be considered as the first step toward more rigorous investigations in such a field of research. Evidently, it needs considerable improvements to be implemented experimentally in future.

Key features of our proposal can be categorized as follows. First, the optical mode with higher damping rate renormalizes both the damping rate and the frequency of the mechanical element (see Appendix \ref{app1}). Second, the mechanical oscillator which operates in the RDR of optomechanics can be adiabatically eliminated to achieve a Kerr-like photon-photon interaction. The description of the system with the NLSE allows us to connect the photonic fluid with the notion of the analog spacetime. We have shown that the phase fluctuation in the photonic fluid obeys the Klein-Gordon equation for a massless scalar field propagating in a curved spacetime with a metric given by Eq.~(\ref{Eq:Metric}) which can be regarded as an analog of the curved spacetime. The corresponding metric can be controlled by the system parameters. 

The system introduced here is rather a simple system. As an outlook for future works, it can be extended to more complex situations which open up several possibilities.  For example, by introducing two primary modes with different polarizations, it should be possible to produce a two-component photonic fluid inside an optical cavity. In principle, it could be possible  to introduce an effective mass for the photons in two dimensions by  adding an optical parametric amplifier in the cavity. Studying the quantum nature of the photon fluid through various quantum optical measurements on the leakage photon field is another outlook for future works.

\appendix
\section{Mechanical oscillator in the RDR} \label{app1}
\begin{figure}
	\begin{center}
		\includegraphics[width=8.5cm]{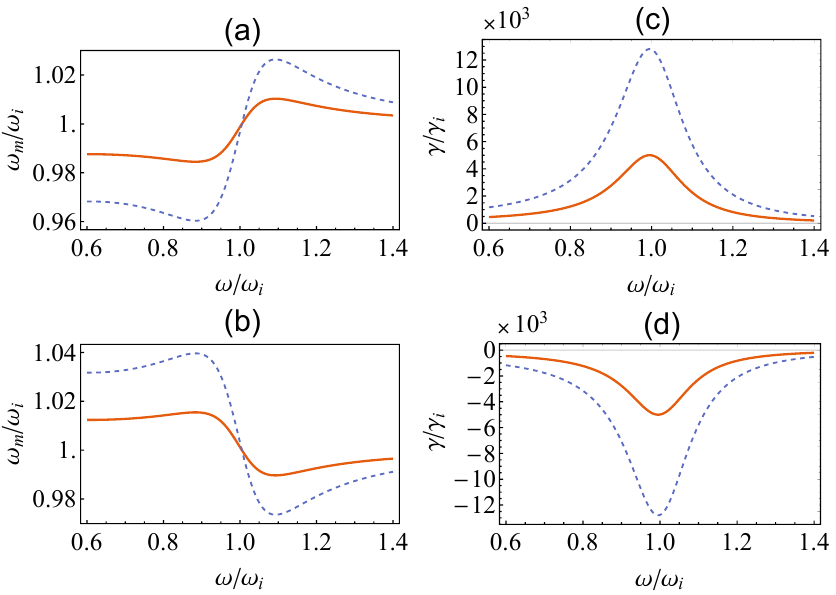}
	\end{center}
	\caption{(color online). (a), (b) Modified frequency of the mechanical oscillator, (c) and (d) modified optomechanical damping rate versus the normalized frequency $\omega/\omega_i$ for two different laser detuning: $\bar{\Delta} = -\omega_i $ (top) and $\bar{\Delta} = \omega_i $ (bottom). The coupling strengths of the ancillary mode are $G = 0.05\omega_i $ (solid line), $G= 0.08 \omega_i$ (dashed line). Parameter values are $\omega_i = 2\pi \times 10 \,\, \rm{MHz}$, $\gamma_i/\omega_i  = 10^{-5}$ and $\kappa' = 0.2\omega_i $. }
	\label{Fig:Fig3}
\end{figure}
In this appendix, we discuss how to prepare the mechanical mode in the RDR of optomechanics. Assume that the mechanical mode with intrinsic frequency $ \omega_i $ is coupled to another ancillary optical mode denoted by $\hat a$ via the radiation pressure. The Hamiltonian of the system in the frame rotating with the laser frequency $ \omega_L $ is given by
\begin{equation}
\hat H/ \hbar={-\Delta \hat a^{\dag}\hat a}+{\omega_i\hat b^{\dag}\hat b}+G_0 \hat a^{\dag}\hat a{({\hat b^{\dag} +\hat b})}+{{(\varepsilon}^* \hat a}+{{\varepsilon}\hat a^{\dag})}\, ,
\end{equation}
where $ G_0 $, $ \varepsilon $ and $ \Delta=\omega_L-\omega_c $ are the single-photon optomechanical coupling, laser pumping rate and the detuning, respectively. Considering a strongly driven field and weak optomechanical coupling allows us to linearize the quantum dynamics of fluctuations around the semiclassical amplitudes, $\hat a = \alpha + \hat c$ and $\hat b = \beta + \hat d$.  The steady-state solutions for the classical mean fields are given by $\alpha  =  - i\varepsilon /\left[ { - \Delta  + G_0(\beta  + {\beta ^*}) - i\kappa' } \right] $ and $ \beta  =  - G_0 {\left| \alpha  \right|^2}/\left( {{\omega _i} - i\gamma_i } \right) $. Here, the cavity damping rate for the ancillary mode is denoted by $ \kappa' $ and the intrinsic damping rate of the mechanical mode is denoted by $ \gamma_i $. The linearized quantum Langevin equations for the fluctuations are obtained as
\begin{equation}
{\partial _t}\hat c =   (i\bar \Delta - \kappa'/2) \hat c -i G\left( {\hat d + {{\hat d}^\dag }} \right) + \sqrt \kappa  \hat c^{in}(t)\,, \label{Eq:LinearizedHeisenbergLangevin1}
\end{equation}
\begin{equation}
{\partial _t}\hat d =   i\left( {{\omega _i} - i{\gamma _i}/2} \right)\hat d - i\left( {G{{\hat c}^\dag } + G^* \hat c} \right) + \sqrt \gamma_i  {{\hat d}^{in}}(t)\,,
\label{Eq:LinearizedHeisenbergLangevin2}
\end{equation}
where we have defined $\bar \Delta  = \Delta  + G_0(\beta + \beta^*)$ and $G=G_0\alpha$. 
The zero-mean operators $\hat{c}^{in}(t)$ and $\hat{d}^{in}(t)$ that denote, respectively, the vacuum optical input noise and the mechanical noise operator satisfy commutation relations $ [ {\hat c^{in}( t),\hat c^{in,\dag }( t')}] = [ {{{\hat d}^{in}}( t),{{\hat d}^{in,\dag }}( t')}] = \delta (t - t')$ and the second order correlations $ \langle {{{\hat d}^{in,\dag }}( t ){{\hat d}^{in}}( t')}\rangle  = {\bar{n}_{th}}\delta (t - t'), $ and $\langle {\hat c^{in}( t)\hat c^{in,\dag }( t')} \rangle  = \delta ( t - t') $ in which we have assumed that the cavity is at zero temperature and $\bar{n}_{th} = {\left( {\exp \left( {\hbar {\omega_m}/{k_B}T} \right) - 1} \right)^{ - 1}}$ is the mean number of thermal phonons of the mechanical oscillator at heat bath temperature $T$ with $k_B$ being the Boltzmann constant. Equations (\ref{Eq:LinearizedHeisenbergLangevin1}) and (\ref{Eq:LinearizedHeisenbergLangevin2}) together with noise correlations fully describe the dynamics of the system under consideration. It is convenient to rewrite the equations of motion in the Fourier space
\begin{equation}
- i\omega \hat c[\omega ] \!=\! ( i\bar \Delta  \!-\! \kappa' /2)\hat c[\omega ] \!-\!i G( {\hat d[\omega ] \!+\! {{\hat d}^\dag }[\omega ]}) \!-\! \sqrt \kappa'  {{\hat a}^{in}}[\omega ],
\end{equation}
\begin{equation}
- i\omega \hat d[\omega ] \!=\!  - \left( i{{\omega _i} \!+\! {\gamma _i}/2} \right)\hat d[\omega ] \!-\! i( {G{{\hat c}^\dag }[\omega ] \!+\! G^*\hat c[\omega ]}) \!-\! \sqrt \gamma_i  {{\hat b}^{in}}[\omega ].
\end{equation}
Combining these two equations results in
\begin{equation}
\hat d[\omega ] = \frac{{ - i\sqrt \kappa'  \left\{ {{G^*}\chi [\omega ]{{\hat a}_{{\rm{in}}}}[\omega ] + G{\chi ^*}[ - \omega ]\hat a_{{\rm{in}}}^\dag [\omega ]} \right\} + \sqrt \gamma_i  {{\hat b}_{{\rm{in}}}}[\omega ]}}{{i\left( {\omega  - {\omega _i} + \Sigma [\omega ]} \right) - {\gamma _i}/2}}\,,
\end{equation}
where $ \chi [\omega ] =  [- i(\omega  + \bar \Delta ) + \kappa' /2]^{-1} $ and $ \Sigma [\omega ] =  - i|G|^2\left( {\chi [\omega ] - {\chi ^*}[ - \omega ]} \right) $ are the optical susceptibility and the self-energy, respectively. Therefore, the effect of the ancillary mode will be a renormalization in the mechanical damping rate ($\gamma_i\rightarrow \gamma$) and the mechanical frequency ($\omega_i\rightarrow\omega_m$). 
The finite cavity lifetime leads to the retarded essence of the radiation-pressure force and, consequently, it introduces the frequency-dependent mechanical frequency shift $\omega _{\rm{opt}}$ and optomechanical damping rate $\gamma _{\rm{opt}}$ given by \cite{Kippenberg,Meystre,Aspelmeyer,Genes,Bowen}
\begin{equation}
{\gamma _{{\rm{opt}}}}(\omega ) = \frac{{{|G|^2}{\omega _i}}}{\omega }\left[ {\frac{{\kappa '}}{{{{\kappa '}^2}/4 + {{\left( {\bar \Delta  + \omega } \right)}^2}}} - \frac{{\kappa '}}{{{{\kappa '}^2}/4 + {{\left( {\bar \Delta  - \omega } \right)}^2}}}} \right],
\end{equation}
\begin{equation}
{\omega _{{\rm{opt}}}}(\omega ) = \frac{{|G|^2{\omega _i}}}{\omega }\left[ {\frac{{\bar \Delta  + \omega }}{{{{\kappa '}^2}/4 + {{(\bar \Delta  + \omega )}^2}}} + \frac{{\bar \Delta  - \omega }}{{{{\kappa '}^2}/4 + {{(\bar \Delta  - \omega )}^2}}}} \right].
\end{equation}
The intrinsic mechanical resonator damping rate and frequency modify due to the radiation pressure as follows
\begin{equation}
\gamma=\gamma_{\rm{opt}}+\gamma_i, \qquad \omega_m=\omega_{\rm{opt}}+\omega_i.
\label{Eq:gammaomega}
\end{equation}
In Fig~(\ref{Fig:Fig3}), we plot the modified frequency of the mechanical oscillator and modified optomechanical damping rate versus the normalized frequency $\omega/\omega_i$ for two different laser detunings $\bar{\Delta} = -\omega_i $ and $\bar{\Delta} = \omega_i $. One can thus increase (by choosing $\bar{\Delta} = -\omega_i $ as shown in Fig~(\ref{Fig:Fig3}-c)) or decrease (by choosing $\bar{\Delta} = \omega_i $ as shown in Fig~(\ref{Fig:Fig3}-d)) the damping rate and frequency of the moving mirror, depending on the sign of the detuning. For a red-detuned ancillary pump there is an increase in the mechanical damping rate of the mechanical oscillator and consequently the cooling of the mechanical oscillator is provided. Moreover, the mechanical oscillator will be spring softened. In  Fig.~(\ref{Fig:Fig3}-a) and Fig.~(\ref{Fig:Fig3}-b) we have plotted, respectively, the modified mechanical frequency and the modified optomechanical damping rate versus the normalized frequency $\omega/\omega_i$ for two different values of the optomechanical coupling of the ancillary optical mode for experimentally feasible parameters of a typical optomechanical system \cite{Vitali,Genes,Aspelmeyer}. The final (steady state) mean phonon number given by  
\begin{equation}
{n}_f = \frac{{{\gamma _{opt}}{{\bar n}_{\min }} + {\gamma _i}{{\bar n}_{th}}}}{{{\gamma _{\rm{opt}}} + {\gamma _i}}}\,,
\end{equation}
where $\bar{n}_{\min}=(\kappa_a/4\omega_{\rm{i}})^2$ is the minimum final phonon number in the resolved-sideband regime \cite{Wilson,Marquardt3}. For the chosen set of parameters in the weak coupling regime $ G < \kappa'/2 $ (normal-mode splitting emerging in the resolved sideband regime at high driving power or strong coupling regime can effect the cooling mechanism \cite{Dobrindt}) at room temperature $ {{\bar n}_{th}}=6.3 \times 10^5$ and $ \bar{n}_{min}=2.5 \times10^{-3} $. The final mean phonon number are $ 4.9\times10^1 $ and $ 1.3\times10^2 $ for  $G = 0.08\omega_i $ and $G= 0.05 \omega_i$, respectively.  Although cooling the mechanical oscillator is very important, it is not necessary to have a small final occupancies (below unity). We should not that with these occupancies the damping rate of the mechanical oscillator can be increased three orders of magnitude. With $\omega=\omega_i$,  and in the resolved side-band limit ($\omega_i \gg\kappa' $) the induced damping rate and the frequency shift of the oscillator are, respectively, given by
\begin{eqnarray}
&&{\gamma _{opt}} = \frac{4|G|^2}{{{\kappa '}}} \,,\label{Eq:DampingRate}\\
&&{\omega _{{\rm{opt}}}} =  - \frac{|G|^2}{{2{\omega _i}}}\,. \label{Eq:FrequencyShift}
\end{eqnarray}
It is evident that $\omega_m$ is negative for $G>\sqrt{2}\omega_i$. Under such a circumstance, the system can enter unstable region and the Routh-Hurwitz stability conditions are violated \cite{Hurwitz}.

Driving the system with an ancillary optical mode with a large damping rate ($\kappa' \gg \kappa$ and $\kappa' \gg \gamma_i$) tuned to the red side of the optical cavity \cite{Marquardt3,Wilson},  as the consequence of the optomechanical interaction, the ancillary mode induces an optical damping to the mechanical oscillator. Working in the resolved-sideband regime and using a  red detuned driving laser together with a small optomechanical single-photon coupling strength which guarantees neglecting the nonlinear optomechanical the net effect of the ancillary optical mode is to renormalize the mechanical damping rate according to Eq.~(\ref{Eq:DampingRate}). 
In this manner, $\gamma_{\rm{opt}}$ can be adjusted by the strength of the driving ancillary field so that the total damping rate of the mechanical oscillator, $\gamma=\gamma _{\rm{opt}}+\gamma _{\rm{i}}$ becomes very large. Therefore, coupling a high-Q mechanical oscillator to an auxiliary cavity mode allows us  to realize the RDR of the cavity optomechanics, $\gamma \gg \kappa$ \cite{Nunnenkamp,Toth} (see Figs. (\ref{Fig:Fig1}-b) and (\ref{Fig:Fig1}-c)). The wide separation between the time scales of dissipation mechanisms allows us to use the mechanical oscillator as an extra dissipative reservoir for the optical mode.

\end{document}